\documentclass[twocolumn]{article}

\usepackage{authblk}
\usepackage{amssymb}

\usepackage{xr}
\usepackage{soul}
\usepackage{color}

\usepackage{amsmath}
\usepackage{amssymb}
\usepackage{caption}
\usepackage{subcaption}
\usepackage{graphicx,float}
\usepackage[export]{adjustbox}
\usepackage{dsfont}
\usepackage{array, booktabs, makecell}
\usepackage[normalem]{ulem}

\usepackage{url}

\begin{document}
\onecolumn
\title{GazeGenie: Enhancing Multi-Line Reading Research with an Innovative User-Friendly Tool}
\date{}

\author[1]{\small Thomas M. Mercier}
\author[2]{\small Marcin Budka}
\author[1,3]{\small Bernhard Angele}
\author[4]{\small Martin R. Vasilev}
\author[1]{\small Timothy J. Slattery}
\author[1]{\small Julie A Kirkby}

\affil[1]{\footnotesize Department of Psychology, Bournemouth University, Poole, Dorset, BH12 5BB, UK}
\affil[2]{\footnotesize Department of Computing \& Informatics, Bournemouth University, Poole, Dorset, BH12 5BB, UK}
\affil[3]{\footnotesize Nebrija Research Centre in Cognition (Centro de Investigación Nebrija en Cognición, CINC), Universidad Antonio de Nebrija, Calle de Asura, 90, 28043 Madrid, Spain}
\affil[4]{\footnotesize Department of Experimental Psychology, University College London, Gower Street, London, WC1E 6BT, UK}

\maketitle

\begin{abstract}
\noindent  In the study of reading, eye-tracking technology offers unique insights into the time-course of how individuals extract information from text. A significant hurdle in using multi-line paragraph stimuli is the need to align eye gaze position with the correct line. This is made more difficult by positional noise in the eye-tracking data, primarily due to vertical drift, and often necessitates manual intervention. Such manual correction is labor-intensive, subjective, and limits the scalability of research efforts. As a result, automated solutions are desirable, especially those that do not require extensive technical skills and still allow close control over the outcome. To address this, we introduce GazeGenie: a comprehensive software solution designed specifically for researchers in eye-tracking studies on multi-line reading. Accessible via an intuitive web browser-based user interface and easily installed using Docker, GazeGenie streamlines the entire data processing pipeline from parsing fixations from raw data to calculation of word and sentence-based measures based on cleaned and drift-corrected fixations. The software's core features include the recently introduced Dual Input Stream Transformer (DIST) model and various classical algorithms all of which can be combined within a Wisdom of the Crowds (WOC) approach to enhance accuracy in fixation line-assignment. By providing an all-in-one solution for researchers, we hope to make automated fixation alignment more accessible, reducing researchers' reliance on manual intervention in vertical fixation alignment. This should lead to more accurate, efficient, and reproducible analyses of multi-line eye-movement data and pave the way to enabling larger scale studies to be carried out. \\

\end{abstract}

\twocolumn
\section{Introduction}

The analysis of reading behavior has emerged as a valuable tool for inferring cognitive processes and proficiency in both natural and programming languages\cite{glandorfSliceAlgorithmAssign2021}. Eye-tracking technology plays a pivotal role in such investigations, allowing researchers to observe participants' eye movements during reading. Millisecond-level recordings of eye movements during reading serve as a reliable source of information for measuring reading behavior, providing insights into the underlying cognitive and motor processes during the processing of single- and multi-line texts, particularly in terms of how the location and timing of gaze fixations reflects ongoing language processing~\cite{raynerEyeMovementsReading1998,rayner35thSirFrederick2009}. ements in eye tracking since the 1970s have contributed to more precise, high-resolution measurements of eye movements and significantly advanced our understanding of reading processes and acquisition~\cite{cohenSoftwareAutomaticCorrection2013}. However, the recorded gaze samples are inherently noisy due to systematic and random eye-tracker error. Even though eye-tracking systems are regularly calibrated throughout an experiment to minimize this error, participants constantly make minute head and body movements leading to calibration drift between and even within trials. Even the initial calibration may be imperfect. This positional eye-tracker error introduces challenges in accurately mapping fixations to units of analysis such as words and sentences~\cite{glandorfSliceAlgorithmAssign2021}. This is particularly true for low-cost eye trackers, which often prioritize accessibility over accuracy. Therefore, the process of line assignment, or mapping fixations to individual lines of a multi-line text is a crucial part of eye-tracking data analysis. Unfortunately, line assignment is a complicated and non-trivial problem in most natural reading settings.

In eye-tracking studies, participants often read single words or sentences, providing valuable data on factors such as the predictability and frequency of target words \cite{rayner35thSirFrederick2009}. However, sentence and paragraph reading experiments are equally crucial in uncovering various levels of written language processing and their cognitive underpinnings, from perceptual span width\cite{blytheVisualInformationCapture2009} to investigations of how word length and frequency influence eye movements\cite{josephWordLengthLanding2009,tiffin-richardsChildrenAdultsParafoveal2015}.

The technical challenge of noise in the tracking data is particularly problematic in multi-line reading studies simply because reading passages takes longer and, therefore, the above-mentioned noise can result in 'vertical drift' - a phenomenon in which fixation recordings inaccurately shift on the vertical axis over time~\cite{limasanchesVerticalErrorCorrection2016}. In other words, fixations might be misplaced above or below the actual line of text participants were reading and these errors may change dynamically as reading progresses. These errors stem from a combination of inherent measurement equipment precision, initial calibration quality, and loss of calibration during the experiment. This complication can hinder data analysis, as fixations are inaccurately mapped to words that were not fixated at that point in time. These errors may be amplified in multi-line reading due to larger text distribution and longer reading time, as well as reduced opportunities for recalibration. Researchers must be mindful of these issues and employ corrective measures to address vertical drift. However, realigning fixations manually can be very time-consuming. Therefore, the creation of specialized software to do this automatically may alleviate some of this burden.

To address the need for automated drift correction and facilitate research in multi-line reading experiments, this work presents GazeGenie an innovative user-friendly tool that does not require any knowledge of programming languages to use effectively. The tool aims to combine data-parsing, data-cleaning, fixation line assignment, calculation of various measures relevant to the analysis and visualization of various aspects of the data in a single software solution. For the necessary task of assigning each fixation to a line of text the user can select from various algorithms, including our recently developed DIST deep learning model~\cite{mercierDualInputStream2024a} and the multi-algorithm WOC framework. This allows for easy comparison between different drift correction methods. GazeGenie offers a user-friendly interface for researchers of varying backgrounds. The resulting tool aims to minimize manual intervention and subjectivity in drift correction processes, ultimately paving the way for a wider-scale adoption of multi-line reading experiments in psychological research. Additionally, it is capable of calculating the most frequently required fixation, saccade, word, and sentence level measures, streamlining data analysis and eliminating the need to switch between various software tools.

The paper's fundamental contribution is the introduction of a robust and reliable tool for reading researchers to undertake larger studies of multi-line reading experiments, enhancing scientific insights and improving efficiency and consistency in data analysis. By streamlining data analysis processes and improving consistency across studies, GazeGenie paves the way for a broader adoption of multi-line reading experiments by researchers.

\section{Related work}
Several software solutions exist for parsing, cleaning, and analyzing eye-tracking data from reading experiments. These include popular options such as popEye~\cite{schroederPopEyePackageAnalyse2022}, Eyekit~\cite{carrEyekit2019}, and EMreading\cite{vasilevMartinvasilevEMreading2023}. Furthermore, there are tools originally developed together with the EyeTrack software at the University of Massachussetts at Amherst\footnote{The software can be found here: \url{https://websites.umass.edu/eyelab/software/}.}: EyeDoctor ~\cite{stracuzziEyeDoctorSoftware2004} and Robodoc, which now largely replaces EyeDoctor, and EyeDry. These tools contributed greatly to making eye-tracking experiments on reading more accessible and standardizing their analysis, but have a somewhat steep learning curve due to their development over many decades, which has led many researchers to develop new solutions like the ones described below which offer more flexibility and may be more accessible.

Each tool has a particular complexity profile, and the intricacies involved in understanding how these tools work can pose challenges, particularly for researchers with less technical backgrounds. This complexity extends not only to the software's core functionalities but also to its integration within research workflows, including data management practices and analytical choices. Each tool comes with its own idiosyncrasies in terms of how it processes and analyzes data.
To put our tool into context we will briefly describe the most popular solutions.

The popEye~\cite{schroederPopEyePackageAnalyse2022} package is an R-based tool designed to analyze eye-tracking data from reading experiments. It allows researchers to analyze data from different experimental paradigms such as single sentence and multi-line reading, boundary change paradigms, and fast priming paradigms using various eye tracking devices like SR Research, SMI, and software packages like EyeTrack or Experiment Builder within the same workflow. It reconstructs stimulus input, parses and cleans data, assigns fixations to linguistic units at different levels, and computes output variables for each level of analysis. The package supports multilingual data processing, including languages like Chinese and Korean, and integrates various automated line alignment algorithms. popEye also calculates measures for sentence-level and text-level processes, providing valuable insights into reading behavior. Additionally, it offers customized plots for various experiment types and can identify problematic trials during data preprocessing. However, it's important to note that at present, popEye only supports data collected using Eyelink eye trackers from SR research and experiments created with SR Research's Experiment Builder or UMass EyeTrack package. Additionally, popEye requires expertise in the R programming language and currently lacks a user interface for novice users. Comparing different data cleaning and line-assignment configurations is also not easily possible, making iteratively dialing in the correct configuration for one's data difficult. To utilize popEye for data analysis, users need ASC files generated by SR Research's EDF2ASC conversion tool, a stimulus file with all experiment items, and an R script with popEye() function calls. This process generates an RDS file containing raw data and aggregated eye movement events on different levels (participant, trial, interest area, etc.) along with specific eye movement measures for each level. These measures can be accessed through generated reports within the RDS file.

EMreading~\cite{vasilevMartinvasilevEMreading2023} is an R package designed to automate the preprocessing and mapping of fixations to text stimuli for single- and multi-line reading experiments recorded with the EyeTrack software and other experimental software that writes experimental data to the ASC file in a similar format. The package aims to provide a user-friendly experience by requiring only a few commands for data preprocessing and also introducing an RShiny-based graphical interface. It offers several functionalities, including data cleaning such as removal of fixations outside the screen or text area, blink removal and handeling of short and overly long fixations. Additionally, users can choose to remove outlier fixations based on standard deviations from the subject's mean. The package also includes functions for calculating standard fixation duration measures at the word level. There are some limitations to consider. First, the package currently only supports data from the EyeTrack software (or other solutions that mimic the way that trials are recorded in EyeTrack), limiting its applicability for researchers using other recording systems. Secondly, it does not integrate automatic line assignment but does allow for correction of the y-position of fixations based on externally computed or human made corrections that are read in separately. Lastly, while a graphical interface is available, users need to have some level of proficiency in R programming to use this package effectively.

The Eyekit~\cite{carrEyekit2019} package is a comprehensive tool for analyzing reading behavior using eye tracking data. The aim is to be hardware-agnostic, making it compatible with various eye tracker systems and data formats. It offers a range of features for users with a background in Python programming, requiring proficiency in the language for its effective utilization. The package employs an object-oriented approach with three key objects: FixationSequence, TextBlock, and InterestArea. FixationSequence handles fixation data, which consists of x-coordinates, y-coordinates, start times, and end times for each fixation. TextBlock represents words, sentences, or text passages and contains details such as positioning and font specifications. InterestArea allows users to define specific areas of interest (words, phrases, morphemes, or letters) in TextBlock objects. These objects interact with one another, enabling users to extract specific portions of texts and compute a range of common reading measures such as gaze duration or initial landing position. Moreover, Eyekit supports various text formats, such as arbitrary fonts, multi-line passages, right-to-left text, and non-alphabetical scripts. It also supports various classical algorithm as well as their combination in a WOC framework to carry out line assignments of fixations~\cite{carrAlgorithmsAutomatedCorrection2022}. Although Eyekit provides an extensive set of tools for eyetracking data analysis, it has some limitations. Its dependency on Python proficiency as a prerequisite might be challenging for some users who are not familiar with the programming language or prefer working with other languages like R or MATLAB. Users who prefer more user-friendly interfaces might also find Eyekit's coding-based approach challenging to navigate initially.

In summary, existing eye-tracking analysis tools like popEye, EMreading, and Eyekit have their own strengths and limitations. While they provide valuable support for researchers in various aspects of eye-tracking data analysis, they often require programming expertise in either R or Python and may not fully address the specific challenge of automated drift correction in multi-line reading experiments. The novel user interface approach presented in our work aims to bridge these gaps by offering data cleaning, automated fixation correction, and data analysis in a user-friendly manner, catering to researchers from diverse backgrounds and paving the way for more efficient multi-line reading experiments and robust scientific insights in human cognition and behavior studies. Another advantage of our tool that sets it apart from prior work is the ability to iteratively determine the configuration for cleaning fixation data and line assignment by adjusting the settings and checking the outcome in automatically generated plots. Our system combines cutting-edge deep learning methodologies and classical algorithms in a collaborative framework to carry out line assignment, ensuring enhanced accuracy and efficiency in eye-tracking data analysis while significantly reducing manual intervention and subjectivity in drift correction processes.

\section{The Program}
Our newly introduced tool offers a user-friendly interface and is designed for researchers of varying backgrounds eliminating the need for extensive programming knowledge in languages like Python, MATLAB or R for researchers wishing to process eye-tracking data from reading experiments. Our software solution consists of a streamlit~\cite{streamlitStreamlitFasterWay2021} based interface and its functionality can be broken down into several key components, which are briefly described below:

\textbf{Data import and preprocessing}: The main focus of GazeGenie is the processing of raw eye-tracking data in ASC format, offering both single file processing and batch processing of multiple files. The program also supports importing single files of custom data in CSV/JSON format, making it possible to use the tool with data from other eye-tracking systems in principle (although the user will have to export their data into the CSV/JSON format; see below). ASC files are the output of one of the most popular eye-trackers, the EyeLink (SR Research, Ottawa, Canada). ASC files are text (ASCII) files  that consist of one row of timestamped coordinates for each new eye sample and additional message lines for identified events such as fixations, saccades,  blinks, as well as other user messages. GazeGenie parses these files to extract all trial fixations, saccades, metadata and stimulus information. Programs such as EyeTrack write the stimulus  coordinate information directly into the ASC file, but other software, such as the proprietary Experiment Builder software (SR Research, Ottawa, Canada) writes the stimulus information into separate interest area files (IAS files). In this case, the ASC file only notes the file name of the IAS file used. In our tool, the IAS files can be added separately and they will be used to align the stimuli coordinates to the correct trials. If the user has their data in a different format and they have already parsed the raw gaze data, the extracted fixations and stimulus files can be imported  CSV or JSON files. The imported data can then be cleaned based on the options selected by the user. Several cleaning options are available: Discarding fixations that are far outside the stimulus text, discarding or merging short fixations and discarding overly long fixations. 

\textbf{Line assignment}: One of the core functions of our software lies in its application of various line assignment algorithms, which includes our DIST deep learning model~\cite{mercierDualInputStream2024a}, classical algorithms\cite{carrAlgorithmsAutomatedCorrection2022} and combinations of both in a WOC framework to accurately map fixations to individual lines in multi-line text to address challenges associated with vertical drift in eye-tracking data. Accurate line assignment is a necessary step for assigning fixations to different parts of the stimulus, such as words and sentences, and therefore underlies any analysis carried out on the data. We have previously shown that our DIST deep learning model is superior to the classical algorithms, especially in a WOC framework~\cite{mercierDualInputStream2024a}. While it is possible to apply the classical algorithms individually using popEye and Eyekit, our tool is the only one that provides a straightforward way to apply the DIST model as well as enabling easy comparison of different line assignment algorithms.

\textbf{Data visualization and analysis}: In analyzing eye-tracking data, it is of great importance to visualize all steps of processing and verify that the chosen parameters for processing and data cleaning are sensible for the stimuli. GazeGenie offers interactive plots and tables for researchers to explore their data and assess the appropriateness of the chosen parsing, cleaning and drift correction configuration. The results of applying one or multiple line assignment algorithms can be visualized by overlaying corrected and uncorrected fixation data on the stimulus material. Based on the corrected fixation data the software offers the option to compute and display various common aggregate fixation, word and sentence measures to support their analysis.

In the following, we will present a guide on how to use the interface and showcase its capabilities.

\subsection{Reading in Data}\label{sec:reading-in-data}
The interface offers two main ways of importing data, depending on whether the user has already extracted the fixations from their experiment or wishes to import raw ASC files. Since information about the bounding boxes of the characters making up the stimulus text is required by the data cleaning procedure, the line assignment algorithms and the subsequent analysis, it is necessary to provide the stimulus information. For preprocessed data, this information may be provided as a CSV or JSON file, whereas for the ASC files, we offer two ways of reading in the stimulus information. Depending on how the eye-tracking experiment is set up, that information may be part of the ASC file, in which case it will automatically be extracted using the "REGION CHAR" flag. In cases where the stimulus information is not included in the ASC file, separate IAS files should be provided containing the coordinates for all words of the texts for each trial. Note that, when using Experiment Builder, these files are required for the experimental setup and should be easily available. In this case, the program expects IAS file name information to be included in the ASC file in lines marked with the "IAREA FILE" flag and other metadata to be marked with the "TRIAL\textunderscore VAR" flag.

\begin{figure}[t]
    \centering
    \includegraphics[width=0.99\linewidth]{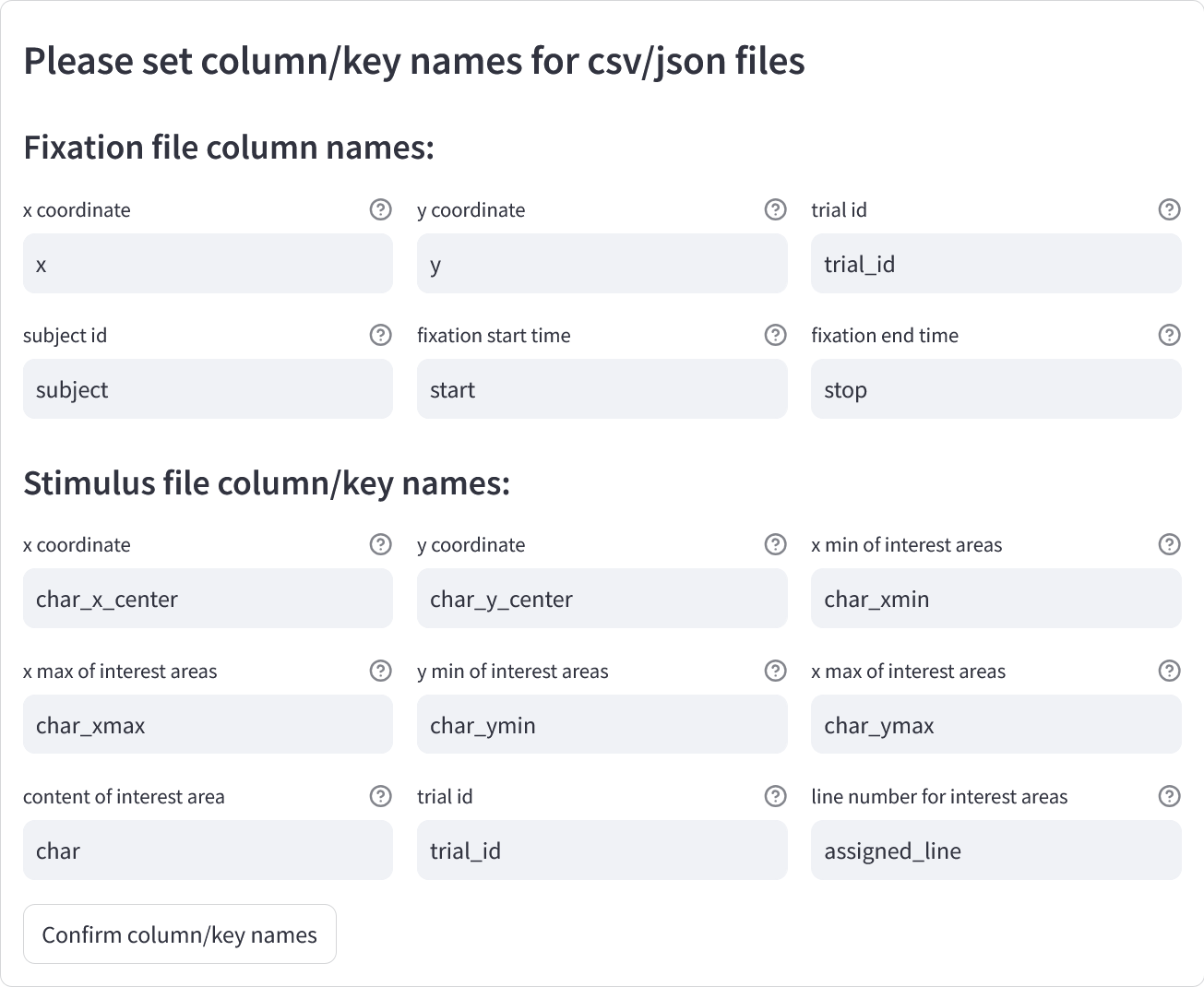}
    \caption{Column and key name inputs fields.}
    \label{fig:colnames_csv}
\end{figure}

For preprocessed fixation data, the user is presented with a preview of the raw loaded data from their files. While GazeGenie will attempt to guess the which of the columns or keys correspond to the required information the user is required to input the column names that they use in their CSV fixation file and the column or key names they use in their CSV or JSON stimulus file if they find the extracted column mappings to be incorrect, as shown in Fig.~\ref{fig:colnames_csv}.

To correctly parse the gaze points, fixations, saccades and blinks for each trial from the ASC file, the user has to choose from a list of flags that mark the start and end of a trial or define custom flags for that purpose. As common starting flags we list "SYNCTIME", "START" and "GAZE TARGET ON". As common stopping flags  we support "ENDBUTTON", "END" and "KEYBOARD". Since it is common for the trial starting event to occur during a fixation, the user can choose to discard that fixation or keep it in the data. In addition the user needs to decide whether spaces between words should be considered part of the word bounding box and if practice and question trial should be excluded when parsing the ASC file. It is recommended to examine the ASC file and choose the start and end flags that occur closest to the onset and offset of the text stimulus. Custom flags (e.g. "DISPLAY ON") can be chosen by selecting the "custom" option and entering the flag in the custom flag text box.

\begin{figure}[t]
	\centering
	\includegraphics[width=0.99\linewidth]{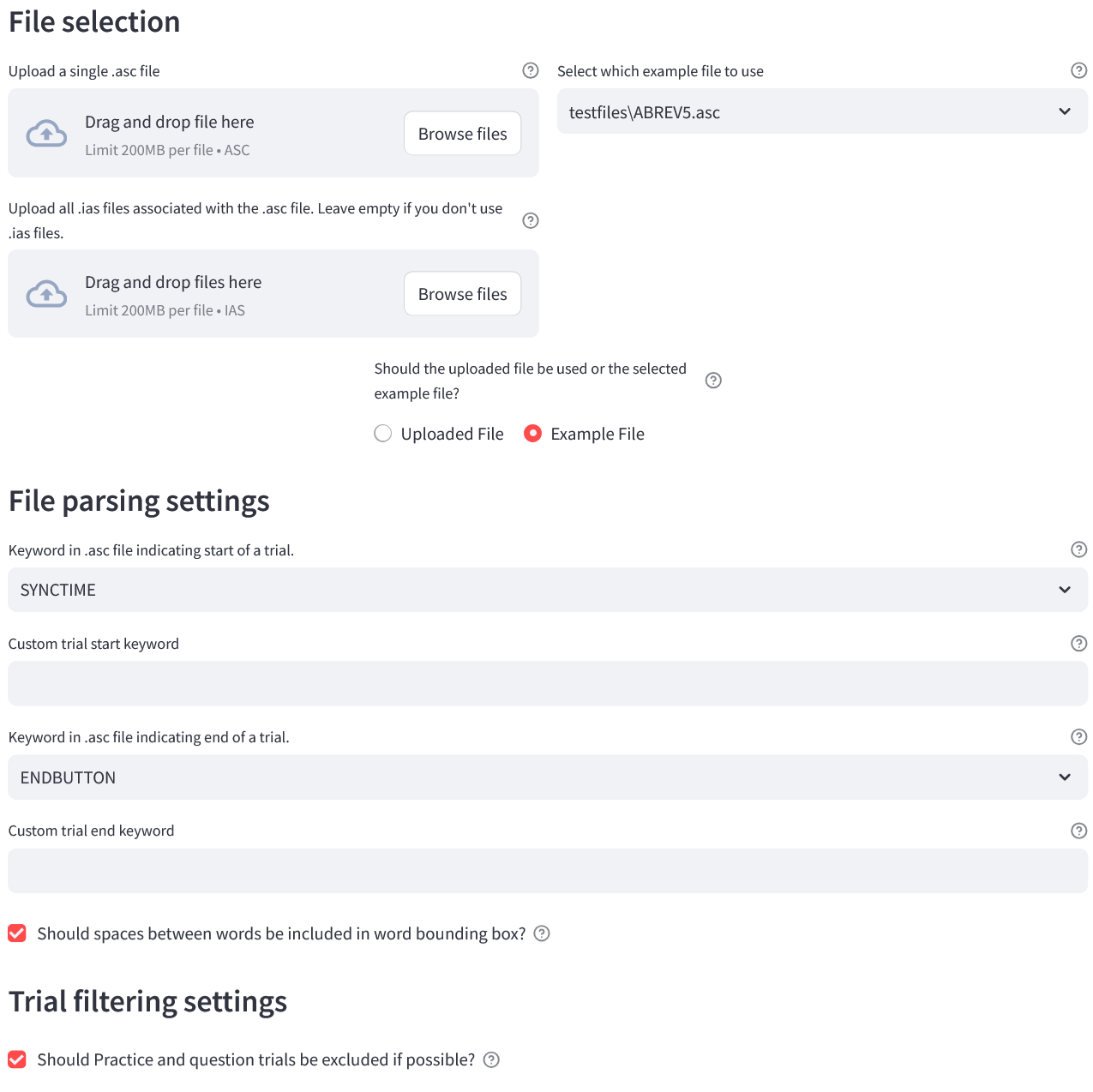}
	\caption{Data reading part of interface for a single ASC file.}
	\label{fig:read_interface}
\end{figure}
Fig.~\ref{fig:read_interface} shows the part of the interface that is used for reading in data. Here the single ASC file part is shown. Some example files are included with the tool to make it easy for a user to familiarize themselves with the tool's functionalities.

The underlying algorithm for parsing ASC files using the provided configuration operates as follows. In the first pass, the algorithm extracts metadata from each trial found in the file, including trial ID, condition, item, question responses, screen size, start and stop indices, timestamps, and character coordinates (if present). For a given trial, the algorithm then iterates over the associated lines to extract gaze points, fixations, saccades, and blinks based on specific flags in the ASC file. Specifically, the algorithm identifies and extracts information about fixations using the EFIX and SFIX flags, blinks using the EBLINK and SBLINK flags, and saccades using the ESACC flag. The extracted data, including gaze points and events, are then further processed to correct start and stop times, calculate event durations, and prepare the data for subsequent analysis or visualization. It is recommended to review the extracted metadata the first time a new data file is read. This can be done by expanding the JSON formatted metadata by clicking on the arrow under "Metadata found in .asc file".

\subsection{Data Cleaning}\label{sec:data-cleaning}

\begin{figure}[t]
	\centering
	\includegraphics[width=0.99\linewidth]{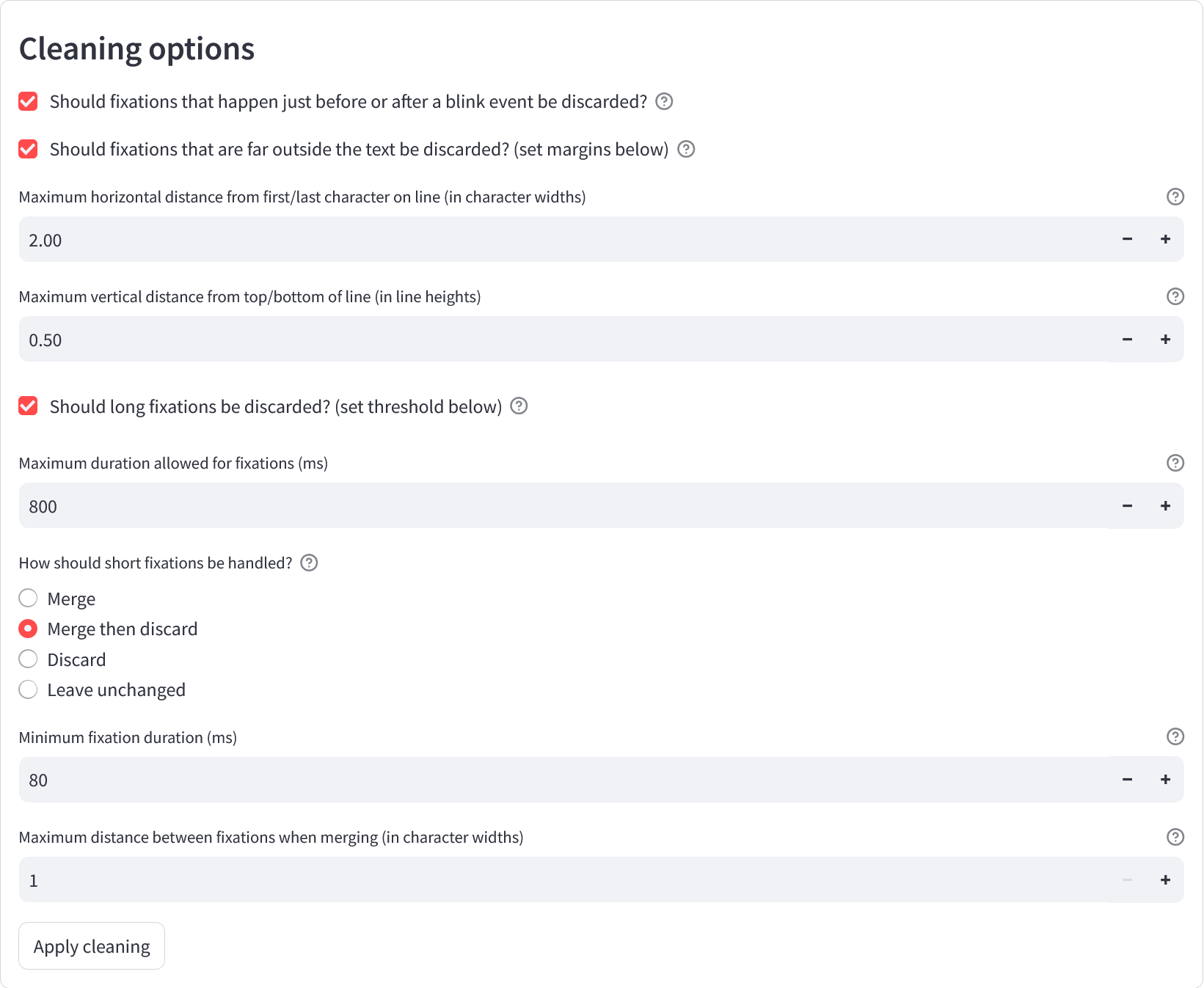}
	\caption{Data cleaning selection for single ASC part of tool.}
	\label{fig:data_clean_single_asc}
\end{figure}

Our tool offers a flexible approach to cleaning fixation data based on duration and position criteria, allowing users to customize the process through various settings. The cleaning options are depicted in Fig.~\ref{fig:data_clean_single_asc}.

Firstly, fixations that occur directly before or after a blink can be discarded. Additionally, users can discard very long fixations by setting a maximum duration (in adults, usually fixations with a duration over 800 or 1000~ms are excluded~\cite{slatteryEyemovementExplorationReturnsweep2019}), and exclude fixations that fall outside all line bounding boxes by adjusting vertical and horizontal distance thresholds. 
To discard fixations that are far outside the text (as defined by the bounding boxes of the lines making up the text), the user sets a horizontal threshold in multiples of the detected character width and a vertical threshold in multiples of line height. For each fixation it is then checked if its coordinates are outside the bounding boxes of all text lines by the required margins.

To address the issue of short fixations, users can choose to merge them with nearby fixations, discard them, merge them where possible and discard the remaining short fixations or leave them unchanged, based on duration and position relative to temporally adjacent fixations. For this the user can set a minimum fixation duration in ms and a maximum distance to adjacent fixations if the merge option is selected. A common minimum fixation duration is 80~ms~\cite{slatteryEyemovementExplorationReturnsweep2019}, which is used as the default setting for this parameter. If either the "Merge" or "Merge then discard" option is selected the cleaning happens as follows. For each fixation with a duration smaller than the threshold, GazeGenie checks if the previous or next fixation can be used to merge. The merging process is contingent on three criteria: the adjacent fixation must not have a blink occurring just before or after it, it must have a duration exceeding the threshold, and the distance between the fixations must be within the specified distance threshold. If these criteria are met, the durations of both fixations are added and the x and y coordinates are averaged. If both the fixation preceding the short fixation and the fixation after it fulfill the merging criteria, then the short fixation will be merged with the longer of the two. If the "Merge then discard" option is selected any remaining short fixations will be discarded. If the "Discard" option is selected all short fixations are filtered out. Default values are provided for all parameters, but users can adjust these settings to suit their specific needs.

\begin{figure}[t]
	\centering
	\includegraphics[width=0.99\linewidth]{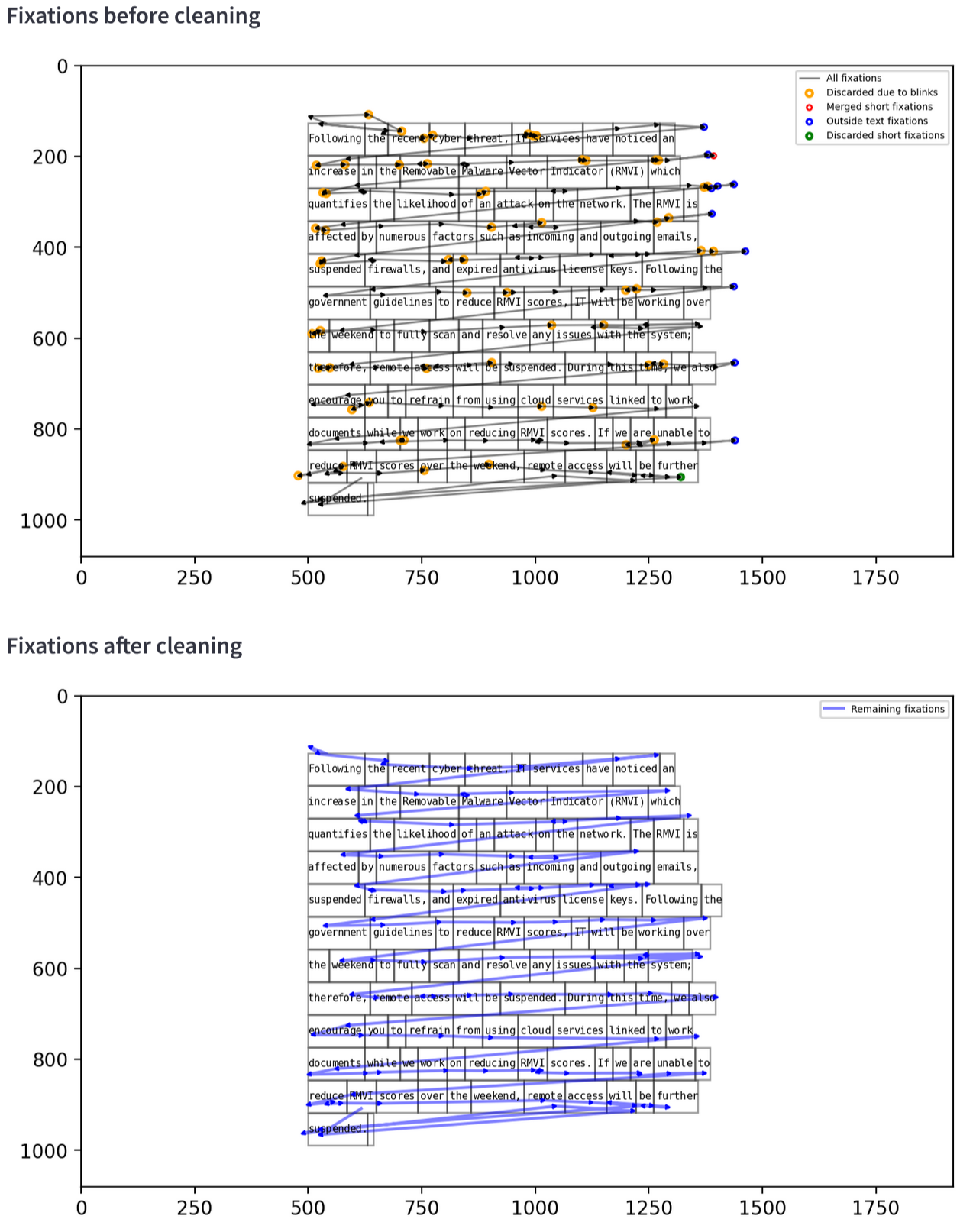}
	\caption{Visualization of data cleaning results.}
	\label{fig:datacleancompare}
\end{figure}
 
Fig.~\ref{fig:datacleancompare} shows how the results of the cleaning process are presented to the user. The top plot highlights which fixations are affected by each part of the cleaning process and the bottom plot shows the fixations remaining after cleaning. In addition, summary statistics are displayed that give an overview of how many fixations were discarded or merged by the cleaning process. The cleaning parameters persist when a different file is loaded or if a different trial is selected for processing and can be saved into a JSON file (see below).

It is important to examine the effects of the data cleaning settings in the visualization generated after running the data cleaning step. This visualization shows the fixations plotted over the stimulus and highlights which fixations were discarded or merged. This way the user can adjust the settings iteratively until they are satisfied with the results. For example, if the out-of-stimulus thresholds are set too narrowly, the visualization will show fixations being discarded that are clearly part of the paragraph reading. In that case, the user can increase the thresholds in order to ensure that all the paragraph reading fixations are included. Note that this is crucial for accurate word level analysis as all remaining fixations will be assigned to a line of text and subsequently to a word on that line. We recommend that users examine multiple trials before deciding on the definitive parameters.

\subsection{Line Assignment}\label{sec:line-assignment}

For successfully parsed and cleaned fixation sequences, GazeGenie offers several drift correction algorithms to assign each fixation to its most appropriate line of text. This results in two columns being added to the fixation data, one containing the line number to which the fixation is assigned and one showing the y-coordinate of the center of corresponding line. Since calculating word or sentence based measures for a trial requires that each fixation is assigned to a line (and subsequently to a word), this capability is the foundation of the analysis features of this tool. 

\begin{figure}[t]
	\centering
	\includegraphics[width=0.99\linewidth]{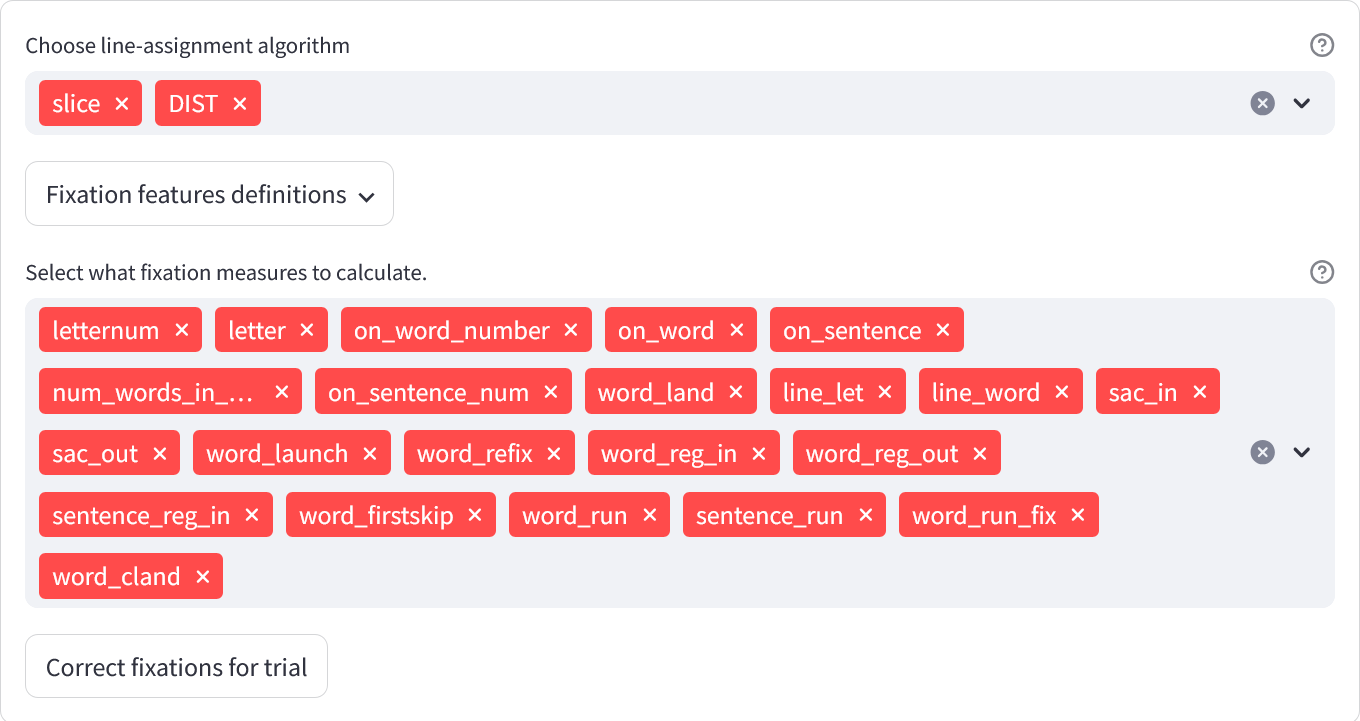}
	\caption{Selection of line assignment approaches and fixation-level features for trial.}
	\label{fig:algochoicesingleasc} 
\end{figure}

The following correction approaches are available: warp, regress, compare, attach, segment, split, stretch, chain, slice, cluster, merge, Wisdom\textunderscore of\textunderscore Crowds, DIST, DIST-Ensemble, Wisdom\textunderscore of\textunderscore Crowds\textunderscore with\textunderscore DIST, Wisdom\textunderscore of\textunderscore Crowds\textunderscore with\textunderscore DIST\textunderscore Ensemble. An overview of the classical approaches can be found in Carr's recent work~\cite{carrAlgorithmsAutomatedCorrection2022} and the deep learning-based approaches are discussed in our recent publication~\cite{mercierDualInputStream2024a}.

As shown in Fig. \ref{fig:algochoicesingleasc}, the user may select one or several of the available correction approaches, with the resulting line assignments and corresponding y-coordinates being marked with the name of the correction approach to allow for easy comparison. At this point, the user may also select which fixation-level features should be calculated.

\begin{figure}[t]
	\centering
	\includegraphics[width=0.99\linewidth]{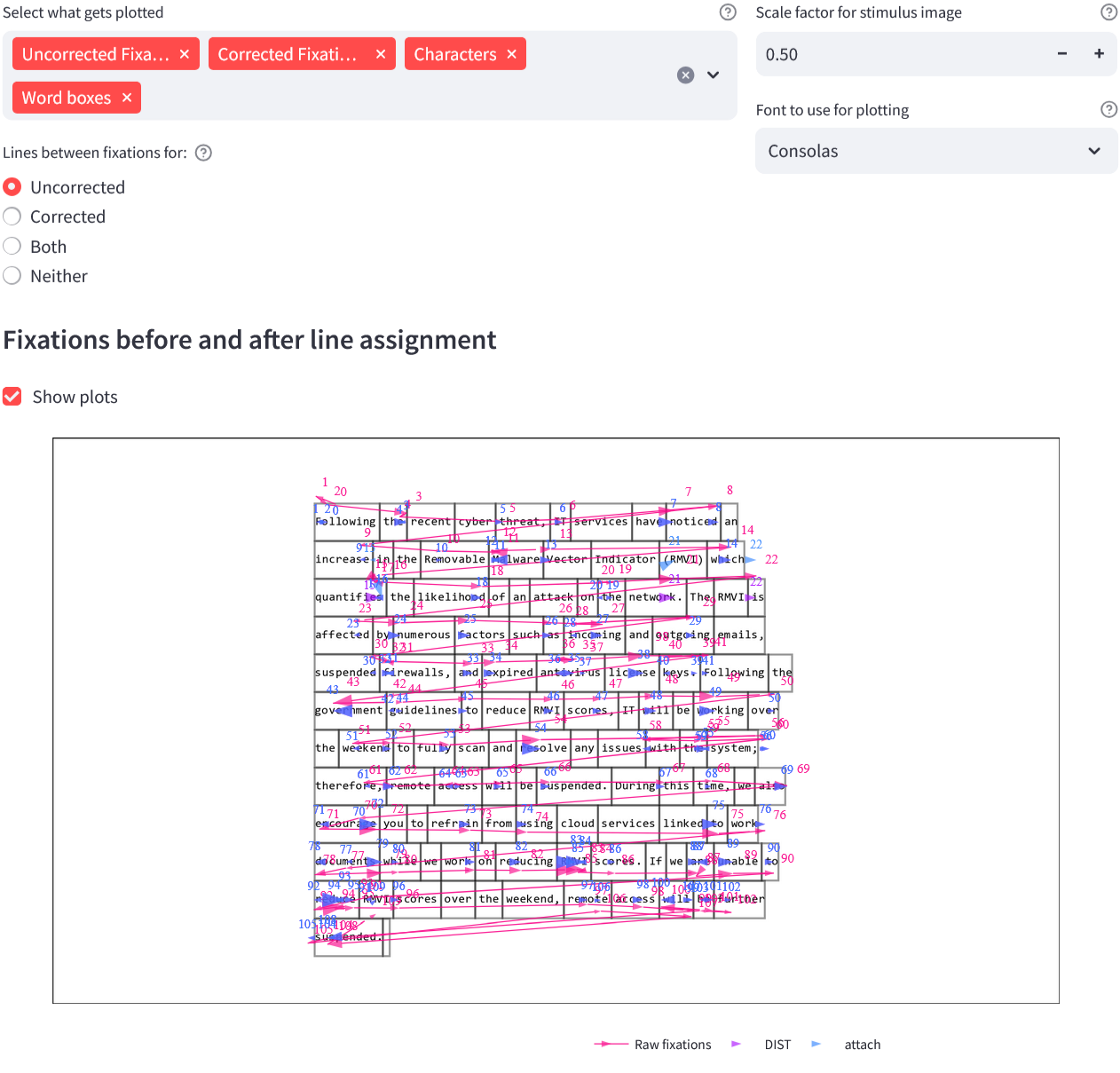}
	\caption{Example of visual comparison of fixation before and after line assignment by DIST and Wisdom of Crowds.} 
	\label{fig:corrfixvrawplot}
\end{figure}

Fig. \ref{fig:corrfixvrawplot} shows how the comparison of the line assignment results is displayed graphically. The plot is interactive such that the results of the selected correction algorithms or the uncorrected fixations may be hidden by the user and to enable closer inspection where needed. Each fixation is numbered to enable easy cross-comparison with the fixation results table. The user also has the option to choose how or if the stimulus is displayed in the plot, the options being:
\begin{enumerate}
	\item Characters - Showing the each character positioned in the center of its bounding box
	\item Character boxes - Showing the bounding boxes for each character
	\item Word boxes - Showing the bounding box for each word, as constructed from all character bounding boxes making up the word
\end{enumerate}
These options can be combined to enable the user to visualize different aspects of the parsed data. 

\begin{figure}[t]
	\centering
	\includegraphics[width=0.99\linewidth]{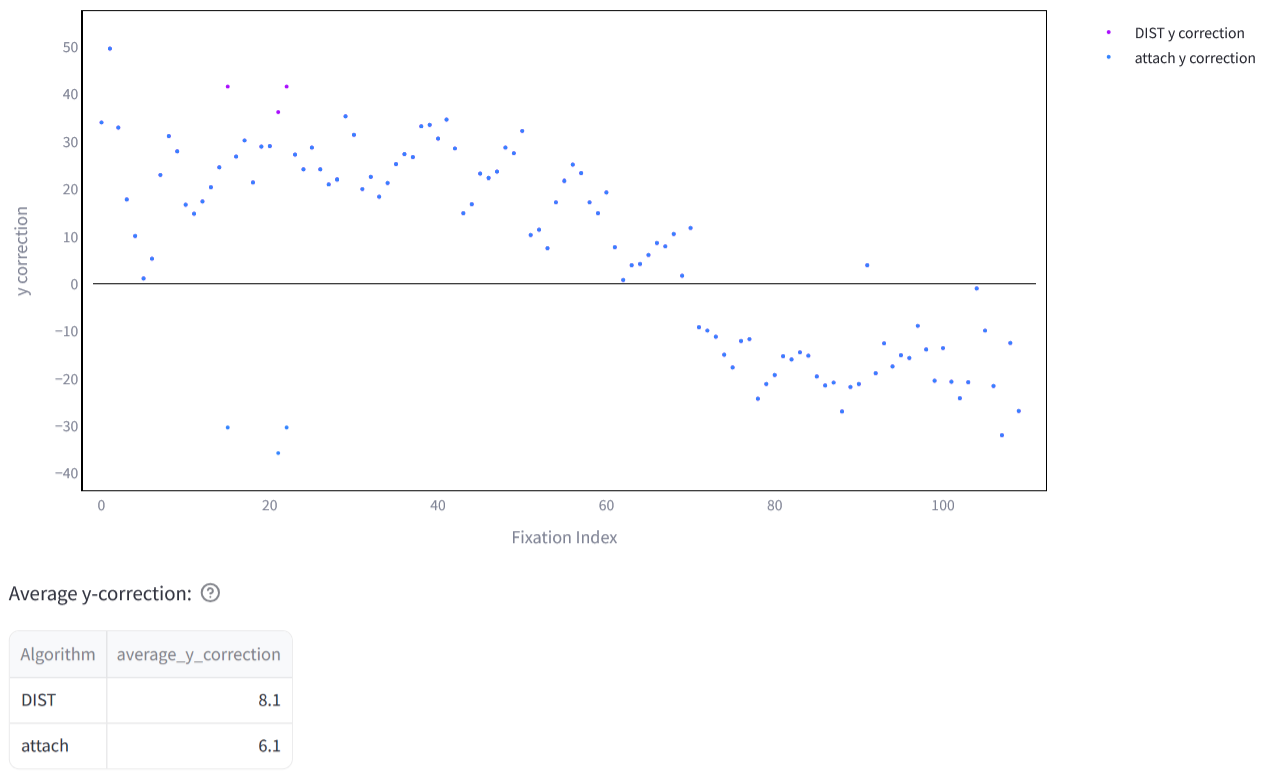}
	\caption{Plot of y-correction in pixels for each selected correction approach.}
	\label{fig:ycorrsingleasc}
\end{figure}

To quickly evaluate the difference between the assignments of the selected algorithms and to spot outliers, the tool also presents a plot of the difference between the uncorrected y-position and the center of the line to which the fixation is assigned in pixels, this is referred to as the y-correction and an example is shown in Fig. \ref{fig:ycorrsingleasc}. An overview of the average y-correction for each algorithm is also presented. We recommend that users examine the output of more than one algorithm, in multiple trials, in order to find the best algorithm for their data (see Section \ref{sec:rec_flow} for recommendations on which algorithms to try in what order).

The cleaning procedure that precedes the line assignment only takes into account the fixations and does not affect the saccades. Because of this, the saccades may no longer be aligned to the fixations. We address this by re-aligning the saccades to the remaining fixations and lines based on temporal alignment with the cleaned fixations. Where saccade features require line information, these features are based on this indirect line assignment.

\begin{figure}[t]
	\centering
	\includegraphics[width=0.99\linewidth]{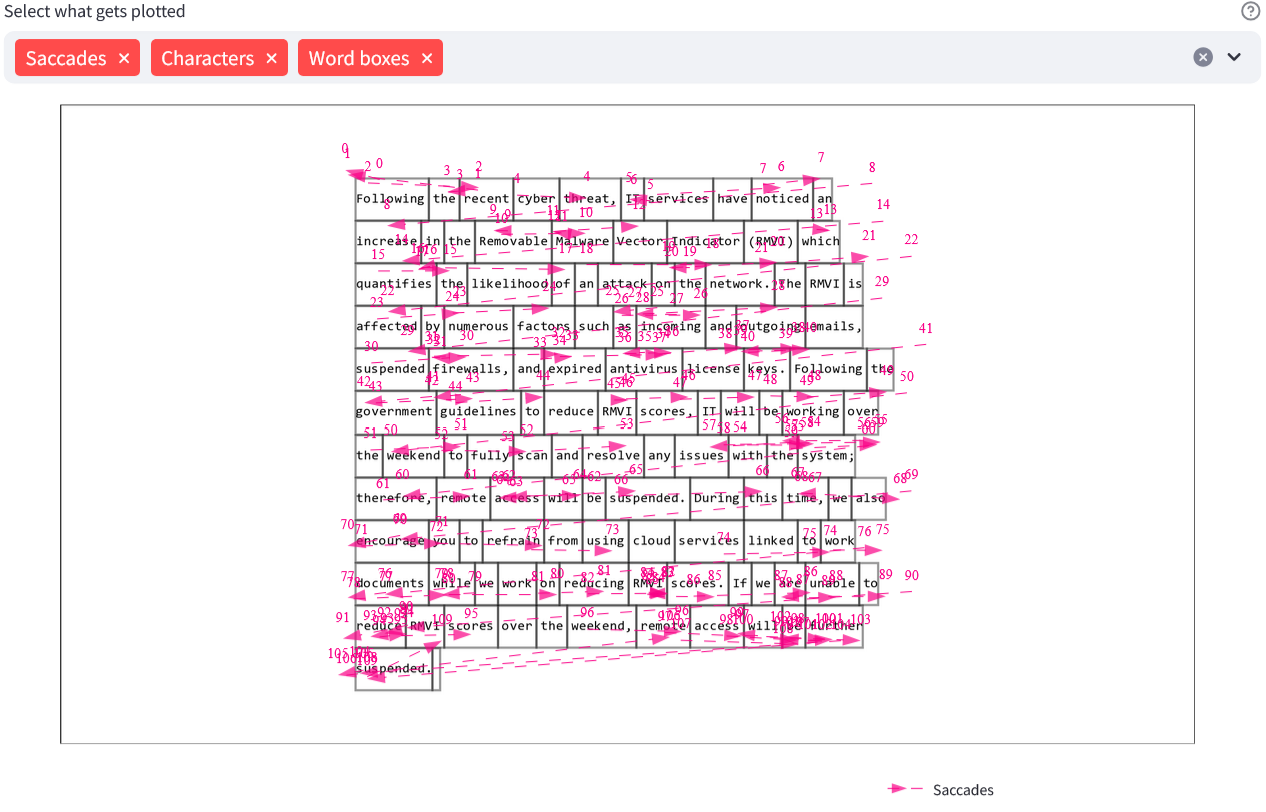}
	\caption{Filtered saccades showing both uncorrected and snapped to lines.}
	\label{fig:saccplotsingleasc}
\end{figure}
The aligned and filtered saccades are displayed in Fig. \ref{fig:saccplotsingleasc}. Both the raw y-coordinates and the snapped to lines coordinates may be displayed.

\subsection{Data Analysis}\label{sec:data-analysis}

Due to the diversity of eye-tracking reading research, we aim to offer a large variety of measures for fixations, saccades, words and sentences. A full description of all these measures is beyond the scope of this publication but can be found in the accompanying documentation (read-me file) of the repository as well as in help messages in the tool itself. Here, we will limit the description to commonly used measures.

For fixations, we calculate common measures such as the landing position relative to the line and within the word, horizontal distance to previous/next fixation in letterwidths, launch distance and whether a blink occurred before or after the fixation. As shown in Fig. \ref{fig:algochoicesingleasc} the user can select which fixation-level measures they would like to be computed.

For saccades we calculate, among others: horizontal saccade length in letter widths, euclidean distance in pixels, angle, whether the saccade constitutes a line change, if it is a return sweep and if it can be considered a directional deviation~\cite{franzenIndividualsDyslexiaUse2021}.

For all downstream word-level measures, it is necessary to assign each fixation to a word. For a given fixation this assignment is done by finding the character closest to it with the search being restricted to the characters on the line of text to which the correction algorithm assigned the fixation. From this it can be calculated which word and which character number relative to the start of the line and word the fixation falls on.

\begin{figure}[t]
	\centering
	\includegraphics[width=0.99\linewidth]{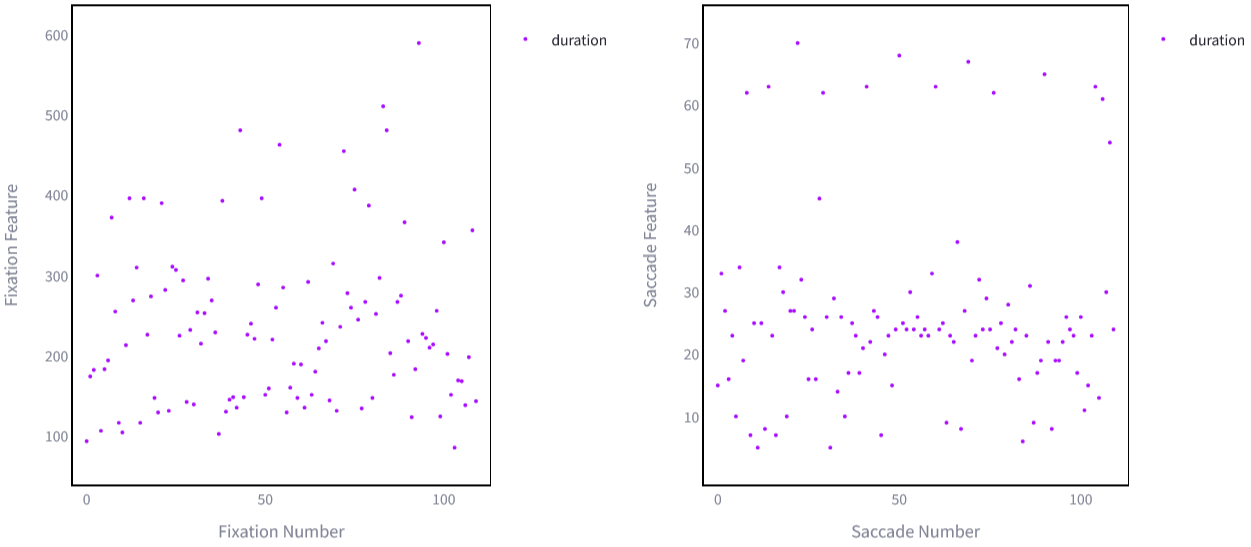}
	\caption{Fixation and saccade feature plots.}
	\label{fig:fixsaccfeatplots}
\end{figure}

Fixation and saccade features can be visualized by selecting the features to plot and whether to use the index or the time as the x-axis, as shown in Fig.~\ref{fig:fixsaccfeatplots}.

For the word level measures we offer, among others:
\begin{enumerate}
	\item first fixation duration \textendash   { }the duration of the fixation on the word during the first-pass
	\item single fixation duration \textendash  { }duration of first fixation on the word
	\item gaze duration \textendash  { }total duration of all fixation assigned to that word
	\item go-past time \textendash  { }duration of all fixations on that word before the word was exited
	\item total number of fixations \textendash  { }total number of fixation that got assigned to the word
\end{enumerate}
 
\begin{figure}[t]
	\centering
	\includegraphics[width=0.99\linewidth]{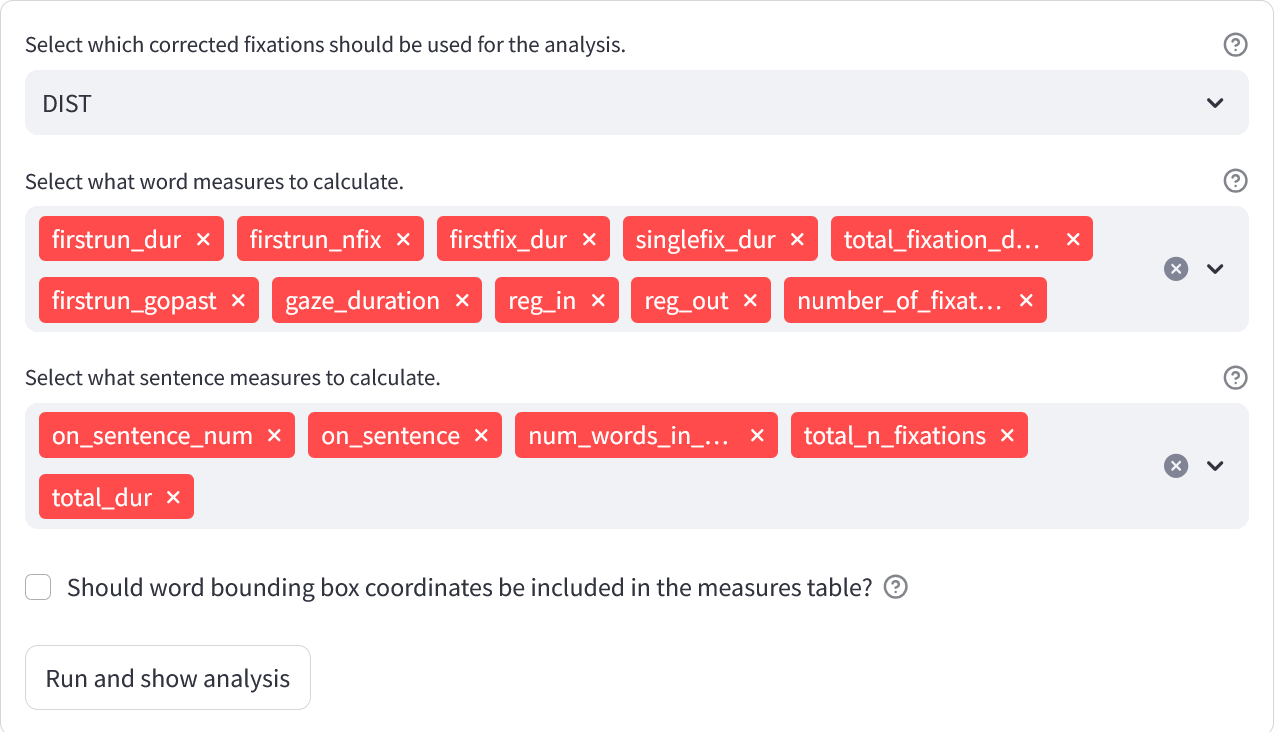}
	\caption{Selection options for word-level measures.}
	\label{fig:word_meas_sel_single_asc}
\end{figure}

The user may select which of the supported word-level measures they wish to compute as shown in Fig.~\ref{fig:word_meas_sel_single_asc}.

\begin{figure}[t]
	\centering
	\includegraphics[width=0.99\linewidth]{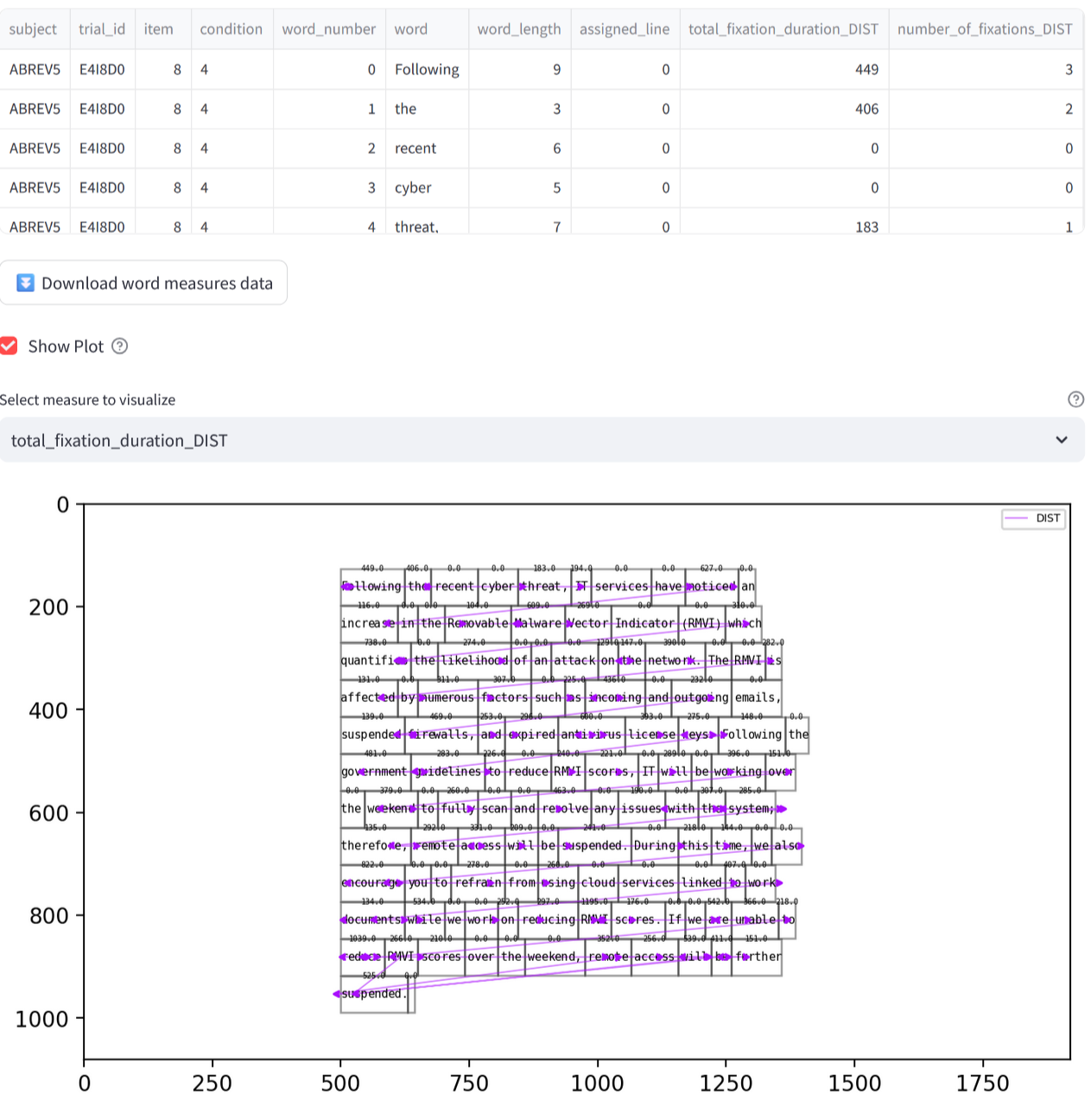}
	\caption{Visualization of corrected fixations and the resulting analysis for total fixation duration for each word.}
	\label{fig:wordanalysissingleasc}
\end{figure}

Fig. \ref{fig:wordanalysissingleasc} shows an example of how the word level features can be visualized with the number at the top of each bounding box giving the value for the selected measure for that word.

\subsection{Batch processing}

In addition to processing individual trials parsed from a single ASC or custom CSV file, the tool offers processing of all trials from one or more ASC files in one step. Once an appropriate configuration has been found by trying different processing settings and correction algorithms and using the visualizations to examine their effects by using the single file tab, the configuration may be saved as a JSON file, which can then be loaded into the batch processing tab. Of course, the settings in the batch processing tab can also be be adjusted manually. To speed up processing of a large number of trials, multiple processor cores can be used to process several of the trials found in an ASC file in parallel. In this case, only a file-level progress bar will be visible during processing. During batch processing, all visualization plots for all trials will be saved and added to a zip file. All resulting data frames for fixations, saccades, words, and sentences will be concatenated and added to the zip file as well. Should the user prefer to save the various outputs for each trial in separate files, this option can also be enabled, but it will result in a large number of files being saved. In addition, summary statistics are calculated, saved, and presented to the user together with the concatenated results, including information about answered questions and fixation cleaning. These data can then be easily imported into statistical software for further analysis. If the user interface is used online, the resulting zip file can be downloaded once processing has finished. For local processing, a results folder will be created in the tool's working directory with the created zip file along with the combined results data frames as CSV files. There is also the option to inspect individual trials from the processed files and plot both the fixation correction results and the analysis similar to how this is done for the single trial processing.

\subsection{Recommended workflow}
\label{sec:rec_flow}
To use GazeGenie effectively, it is recommended to follow these steps. Given a set of ASC files as the starting point, the user should first load one file via the single file tab. Depending on how the reading experiment was set up initially, additional IAS files containing the stimulus information may be required and the flags indicating the start and end timestamp of a trial might need to be adjusted. After loading in the ASC file, the tool does a first parse and the metadata for all trials found in the file are displayed and can be checked by the user. If the extracted metadata and trial overview are as expected, a single trial should be selected, loaded, and the resulting fixation, saccade and blink data should be inspected. The next step is the cleaning of the fixation data. To find appropriate cleaning parameters it is recommended to start with the given default configuration, this would mean that fixations longer than 800~ms and fixations that occur just before or just after a blink event would be discarded, fixations shorter than 80~ms would be merged into the previous or subsequent fixation as long as they are one character width or less from each other. Short fixations that could not be merged would then be discarded. It is recommended to try several combinations with a range of parameter values and inspect the resulting plot of which fixations are affected by the cleaning process and the summary cleaning statistics each time and adjust the parameters accordingly. Vertical and horizontal thresholds for excluding fixations outside the stimulus area should be paid particular attention to as they can easily result in a large number of discarded fixations. These thresholds should only exclude fixations that are clearly not part of the reading process. Since all remaining fixations will be assigned to a word, this can have a significant effect on the downstream analysis. This step should be repeated for several trials to make sure the settings are appropriate. Once the cleaning parameters have been adjusted, one or more correction algorithms need to be selected. Depending on the level of drift present in the data, a simple classical algorithm may be sufficient and will be advantageous in terms of processing speed. Slice, regress and merge are good choices to start with. Should the classically achieved results be unsatisfactory, it is recommended to use the DIST algorithm and use the WOC + E-DIST approach if necessary. Here again it is recommended to visually inspect the resulting plots to gauge which correction approach is best suited for the data. Using the results from applying the correction algorithm, the user then chooses which word measures should be calculated based on the line assignments from the correction step. Here again, the results can be inspected by both examining the word and sentence measures tables directly or by visualizing the word measures. Once the user is satisfied with the configuration it can be saved as a JSON file to be read in again at a later time or to be used for the batch processing tab. With an appropriate configuration for the batch processing the user can either load in a single ASC file and process all trials or add all their ASC files to process everything in one step. To increase the speed with which files can be uploaded into the batch processing tab, GazeGenie supports adding files via a zip file. The results can then be inspected by viewing all combined tables or the different plots.

\subsection{Accessing GazeGenie}

The tool can be accessed through multiple avenues, catering to diverse user needs and preferences. For setup-free access, a hosted online version is available via Huggingface Spaces, which eliminates the need for local installation and configuration. However, this approach limits the available computational power and requires uploading data to Huggingface servers and subsequently downloading processed results, which can take a significant amount of time.

Local installation offers distinct advantages, including accelerated data reading and automatic saving of results, which can significantly enhance user experience and workflow efficiency. There are two ways of using GazeGenie locally. The recommended way is to use the Docker image, which only requires Docker to be installed on the local system. After that, the tool can be run by simply pulling and running the docker image. The exact commands to run the tool via docker will be found in the repository associated with this article. This way of installing the tool is recommended as it requires the least amount of technical knowledge. The second option requires some technical knowledge and is mainly interesting for expert users who wish to modify GazeGenie themselves. In this case, users need to set up Python and all required packages for a local installation. For this, it is recommended to utilize the Miniforge distribution to install Python and the necessary packages, ensuring a seamless setup process. Once Python is installed, users clone the repository from GitHub~\footnote{Github repository: \url{https://github.com/Gittingthehubbing/GazeGenie}}, after which the requisite packages can be installed via conda or pip. Notably, using pip with the provided requirements file in the repository offers a more convenient option, as it enables direct downloading and installation of all required packages.

The tool's interface is built upon the Streamlit package, and thus requires initiation via the corresponding Streamlit command. Detailed instructions for execution can be found in the accompanying README file. Upon launching the tool, an interactive interface will open in the browser, allowing users to interact with it effortlessly.

\subsection{Limitations of the tool}
Like any solution, ours too has its limitations. All of the data available for testing and developing GazeGenie was based on data from the most popular Eyetracking hardware, the Eyelink. While our tool does support reading in custom CSV files, this does rely on the user checking that the column names are mapped correctly and for them to make adjustments if needed. For custom files, it can also only read in data in the form of extracted fixations and does not offer reading in raw gaze data with subsequent automatic fixation extraction. We also acknowledge that reading research using eye-tracking is a diverse field and our tool might not work well for experiments that are very different from a simple paragraph reading setup.

\section{Conclusion}
We have introduced GazeGenie, a novel tool to process eye-tracking data from passage reading experiments. Our tool's data parsing, cleaning, correction and analysis functionalities, alongside with its visualization features, streamline the post-processing steps necessary to extract meaningful insights from eye-tracking datasets and will facilitate more accurate and reproducible research outcomes. Importantly, our tool allows even users with limited technical skills to use our DIST model (along with all other fixation alignment algorithms) for aligning fixations to lines in multi-line stimuli. This is a significant step towards making the analysis of multi-line data more efficient and accessible to a larger number of researchers. 

\section{Declarations}

\subsection{Funding}

This work was made possible in part by a research grant from the Microsoft Corporation to T.J.S. which was matched by Bournemouth University to fund PhD studentships.

\subsection{Conflicts of interest}
The authors have no relevant financial or non-financial interests to disclose.

\subsection{Ethics approval}
Not applicable

\subsection{Consent to participate}
Not applicable

\subsection{Consent for publication}
Not applicable

\subsection{Replication}

Not applicable

\subsection{Availability of data and materials }

Not applicable

\subsection{Code availability}

The code is available at: \url{https://github.com/Gittingthehubbing/GazeGenie}

\subsection{Open Practices Statement}

No experiments have been presented, therefor preregistration is not applicable.

The data and materials/code are available at: \url{https://github.com/Gittingthehubbing/GazeGenie}
\section{Pre-print note}
This work has been submitted to the Springer for possible publication. Copyright may be transferred without notice, after which this version may no longer be accessible. Personal use of this material is permitted. Permission from Springer must be obtained for all other uses, in any current or future media, including reprinting/republishing this material for advertising or promotional purposes, creating new collective works, for resale or redistribution to servers or lists, or reuse of any copyrighted component of this work in other works.
\bibliographystyle{abbrv}
\bibliography{Eye-Tracking}
\end{document}